\documentclass[preprint,prl,10pt,twocolumn]{revtex4}%
\usepackage{amsfonts}
\usepackage{amsmath}
\usepackage{amssymb}
\usepackage{graphicx}%
\begin{document}
\title{Enhancing Monte Carlo methods by using a generalized fluctuation theory}
\author{L. Velazquez}
\affiliation{Departamento de F\'{\i}sica, Universidad de Pinar del R\'{\i}o, Mart\'{\i}
270, Esq. 27 de Noviembre, Pinar del R\'{\i}o, Cuba.}

\begin{abstract}
According to the recently obtained \textit{thermodynamic uncertainty relation}
$\left\langle \delta\beta\delta E\right\rangle =1+\left(  \partial
^{2}S/\partial E^{2}\right)  \left\langle \delta E^{2}\right\rangle $, the
microcanonical regions with a negative heat capacity can be accessed within a
canonical-like description by using a thermostat with a fluctuating inverse
temperature. This far-reaching conclusion is used in this Letter for enhancing
the potentialities of the well-known Swendsen-Wang cluster algorithm in order
to access to the anomalous microcanonical states of the $q=10$ state Potts
model on a square lattice $L\times L$, which exhibits a first-order phase
transition in its thermodynamical description. 

\end{abstract}
\maketitle

The present approach for enhancing the Monte Carlo methods based on the
equilibrium distributions of the Statistical Mechanics \cite{mc3} is
straightforwardly followed as a direct application of a non Riemannian
framework of the Equilibrium Thermodynamics recently proposed in the
refs.\cite{vel.geo,vel-mmc}. A fundamental result derived from this theory is
the following thermodynamic identity:%
\begin{equation}
\left\langle \delta\beta\delta E\right\rangle =1+\frac{\partial^{2}S}{\partial
E^{2}}\left\langle \delta E^{2}\right\rangle ,\label{tur}%
\end{equation}
where $\beta$ characterized the \textit{effective inverse temperature} of
certain generalized thermostat and $E$ the total energy of the interest
system. This expression generalizes the well-known relation $C=\beta
^{2}\left\langle \delta E^{2}\right\rangle $ between the heat capacity
$C=dE/dT$ and the energy fluctuations within the Gibbs canonical description
towards all those microcanonical regions hidden by the ensemble inequivalence,
that is, all those energetic regions characterized by the presence of
\textit{negative heat capacities} \cite{gro1}:%
\begin{equation}
C=-\left(  \frac{\partial S}{\partial E}\right)  ^{2}\left(  \frac
{\partial^{2}S}{\partial E^{2}}\right)  ^{-1},\label{c.def}%
\end{equation}
associated to the convex character of the Boltzmann entropy $S=\ln W$.

The thermodynamic identity (\ref{tur}) naturally appears within the
generalized framework under the hypothesis of the \textit{ensemble
equivalence}. Thus, the ordinary condition $\delta\beta=0$ associated to the
Gibbs canonical description is only admissible wherever the entropy be a
concave function $\partial^{2}S/\partial E^{2}<0$ (convex-down). The identity
(\ref{tur}) leads to the following inequality:%
\begin{equation}
\left\langle \delta\beta\delta E\right\rangle \geq1\label{ine}%
\end{equation}
wherever the entropy be locally convex $\partial^{2}S/\partial E^{2}\geq0$
(convex-up). This means that the anomalous regions with a negative heat
capacity could be only accessed by using a generalized thermostat exhibiting
correlated fluctuations of its effective inverse temperature $\beta$ with the
fluctuations of the total energy $E$ of the interest system. This far-reaching
conclusion could be a fundamental ingredient for allowing the improvement of
many Monte Carlo algorithms based on the Gibbs canonical ensemble for deal
with the presence of first-order phase transitions without the necessity of
appealing to multicanonical methods \cite{mc3}. The aim of the present Letter
is to show a numerical evidence which seems to support the above ideas.

The reader can also notice that the inequality (\ref{ine}) looks-like an
uncertainty relation, which allows us to claim that the inverse temperature of
the thermostat $\beta$ and the total energy of the interest system $E$ can be
considered as complementary thermodynamical quantities within the regions of
ensemble inequivalence. The energy fluctuations $\delta E$ can not be reduced
there without an increasing the fluctuations of the inverse temperature
$\delta\beta$ of the generalized thermostat, and vice versa. Thus, the
identity (\ref{tur}) can be referred as a \textit{thermodynamic uncertainty
relation} since it imposes certain restrictions to the determination of the
microcanonical caloric curve.

The understanding of the methodology introduced in this Letter demands to
carry out a summary of the geometric foundations leading to the thermodynamic
uncertainly relation (\ref{tur}). I shall only exposed here the fundamental
points, so that, the interested reader should see the
refs.\cite{vel.geo,vel-mmc} for more details.

Let be an isolated Hamiltonian system $\hat{H}_{N}$ whose macroscopic
description can be performed starting from microcanonical basis by considering
only the total energy $E=\hat{H}_{N}$. Let $\Theta=\Theta\left(  E\right)  $
be any diffeomorfism (a bijective and piece-wise differentiable function) of
the total energy $E$. It is said that the functional $\hat{\Theta}_{N}%
=\Theta\left(  \hat{H}_{N}\right)  $ constitutes a reparametrization of the
Hamiltonian $\hat{H}_{N}$. It is easy to show that the Physics within the
microcanonical description is\textit{ reparametrization invariant}:%
\begin{align}
\hat{\omega}_{M}\left(  E\right)   &  =\frac{1}{\Omega\left(  E,N\right)
}\delta\left[  E-\hat{H}_{N}\right]  ,\nonumber\\
&  \equiv\frac{1}{\Omega\left(  \Theta,N\right)  }\delta\left[  \Theta
-\hat{\Theta}_{N}\right]  =\hat{\omega}_{M}\left(  \Theta\right)
.\label{rep.inv}%
\end{align}
The demonstration is straightforwardly followed from the properties of the
Dirac function. Such feature implies that the microcanonical average
$\left\langle A\right\rangle =Sp\left(  \hat{\omega}_{N}\hat{A}\right)  $ of
any microscopic observable $\hat{A}$ is also reparametrization invariant,
$\left\langle A\right\rangle \left(  E\right)  =\left\langle A\right\rangle
\left(  \Theta\right)  $, that is, these averages can be taken as scalar
functions under the energy reparametrization changes. The microcanonical
partition function $\Omega\left(  \Theta,N\right)  =Sp\left\{  \delta\left[
\Theta-\hat{\Theta}_{N}\right]  \right\}  $ allows us to introduce the
invariant measure of the phase space volume:
\begin{equation}
d\mu=\Omega\left(  \Theta,N\right)  d\Theta=\Omega\left(  E,N\right)
dE,\label{measure}%
\end{equation}
which leads to an reparametrization invariant definition of the microcanonical
entropy $S=\ln W_{\alpha}$, where $W_{\alpha}=\int_{\Sigma_{\alpha}}d\mu$,
being $\Sigma_{\alpha}$ a subset of certain coarsed grained partition of the
phase space. Such a coarsed grained nature of the entropy can be disregarded
in the thermodynamic limit $N\rightarrow\infty$ and this thermodynamic
potential can be considered as a \textit{continuous scalar function}. The
estimation $W_{\alpha}\simeq\Omega\left(  \Theta,N\right)  \delta\Theta_{0}$
can be used whenever the subset $\Sigma_{\alpha}$ be small, a condition which
is always justified when $N$ is large enough without any lost of generality
\cite{vel.geo}.

The concavity of a scalar function as the microcanonical entropy $S$ depends
crucially on the reparametrization used for describing the thermodynamical
properties, which is explained by the presence of the term $\partial
S/\partial\Theta$ in the transformation rule of the entropy Hessian during the
reparametrization $E\leftrightarrow\Theta$:
\begin{equation}
\frac{\partial^{2}S}{\partial E^{2}}=\left(  \frac{\partial\Theta}{\partial
E}\right)  ^{2}\frac{\partial^{2}S}{\partial\Theta^{2}}+\frac{\partial
^{2}\Theta}{\partial E^{2}}\frac{\partial S}{\partial\Theta}. \label{transf}%
\end{equation}
A trivial example is the concave function $s\left(  x\right)  =\sqrt{x}$ with
$x>0$, which becomes a convex function $s\left(  y\right)  =y^{2}$ after
consider the reparametrization $y=\sqrt[4]{x}$.

The concavity of the microcanonical entropy is an important condition for the
ensemble equivalence. The modifying of the entropy convex properties during
the energy reparametrization changes is a very simple alternative for ensure
the ensemble equivalence within a canonical-like description. This kind of
framework demands the introduction of the called \textit{generalized Gibbs
canonical ensemble}:%
\begin{equation}
\hat{\omega}_{GC}\left(  \eta\right)  =\frac{1}{Z\left(  \eta,N\right)  }%
\exp\left[  -\eta\hat{\Theta}_{N}\right]  ,\label{gce}%
\end{equation}
which describes the equilibrium conditions of the interest system under the
external influence of a generalized thermostat with generalized canonical
parameter $\eta$. It is easy to show that this distribution function is
derived within the Jaynes reinterpretation of the Thermodynamics in terms of
the Information theory \cite{jaynes} by imposing the constrain:%
\begin{equation}
\left\langle \Theta\right\rangle =\sum_{k}\Theta\left(  E_{k}\right)
p_{k},\label{constrain}%
\end{equation}
instead of the usual energy constrain with $\Theta\left(  E\right)  =E$, a
viewpoint recently developed by Toral in the ref.\cite{toral1}. This last
observation clarifies us that the role of the generalized thermostat is
precisely to impose the constrain (\ref{constrain}).

The generalized partition function $Z\left(  \eta,N\right)  =Sp\left\{
\exp\left[  -\eta\hat{\Theta}_{N}\right]  \right\}  $ can be rewritten in
terms of the microcanonical partition function $\Omega\left(  E,N\right)  $ as
follows:
\begin{align}
Z\left(  \eta,N\right)   &  =\int\exp\left[  -\eta\Theta\right]  \Omega\left(
E,N\right)  dE\nonumber\\
&  \equiv\int\exp\left[  -\eta\Theta\right]  \Omega\left(  \Theta,N\right)
d\Theta,
\end{align}
where the invariant character of the measure (\ref{measure}) possibilities us
to rephrase the generalized partition function as an usual \textit{Laplace
transformation}. The imposition of the thermodynamic limit $N\rightarrow
\infty$ leads to the validity of the \textit{Legendre transformation} between
the Thermodynamic potentials $P$ and $S$:%
\begin{equation}
P\left(  \eta\right)  =\inf_{\Theta^{\ast}}\left\{  \eta\Theta-S\left(
\Theta,N\right)  \right\}  ,
\end{equation}
being $P\left(  \eta,N\right)  =-\ln Z\left(  \eta,N\right)  $ the Planck
potential associated to the generalized canonical ensemble (\ref{gce}),
whenever there is only one point $\Theta^{\ast}$ satisfying the stationary
conditions:%
\begin{equation}
\eta=\frac{\partial S\left(  \Theta^{\ast},N\right)  }{\partial\Theta}\text{
and }\kappa_{\Theta}=\frac{\partial^{2}S\left(  \Theta^{\ast},N\right)
}{\partial\Theta^{2}}<0\text{.} \label{station}%
\end{equation}
within the energy reparametrization $\Theta=\Theta\left(  E\right)  $.

While the microcanonical ensemble is reparametrization invariant
(\ref{rep.inv}), the generalized Gibbs canonical description (\ref{gce})
depends crucially on the energy reparametrization. Nevertheless, these
equilibrium statistical descriptions becomes asymptotic equivalent with the
imposition of the thermodynamic limit $N\rightarrow\infty$ with the exception
of all those microcanonical regions where the convex-down character of the
entropy $S$ in the energy reparametrization $\Theta$ can not be ensured.

The reader can notice that the using of energy reparametrizations
$\Theta=\Theta\left(  E\right)  $\ allows us to extend many results and
methodologies of the standard Thermodynamics and Boltzmann-Gibbs Statistical
Mechanics with a simple change of reparametrization $\left(  \beta,E\right)
\Rightarrow\left(  \eta,\Theta\right)  $. This viewpoint was considered in the
ref.\cite{vel-mmc} for introduce a variant of the Metropolis importance sample
algorithm \cite{metro} based on the generalized canonical weight $\exp\left[
-\eta\Delta\Theta\right]  $ instead on the usual $\exp\left(  -\beta\Delta
E\right)  $, being $\Delta\Theta=\Theta\left(  E+\Delta E\right)
-\Theta\left(  E\right)  $. Since $\left\vert \Delta E\right\vert <<\left\vert
E\right\vert $ during a Metropolis move when $N$ is large enough,\ the
approximation $\Delta\Theta\simeq\partial\Theta\left(  E\right)  /\partial
E\ast\Delta E$ allows us to rephrase the corresponding acceptance probability
$p$:%
\begin{equation}
p\simeq\min\left\{  1,\exp\left[  -\hat{\beta}\Delta E\right]  \right\}  ,
\end{equation}
as an ordinary Metropolis move with a \textit{variable inverse temperature}:%
\begin{equation}
\hat{\beta}=\beta\left(  E;\eta\right)  =\eta\frac{\partial\Theta\left(
E\right)  }{\partial E}. \label{variable.b}%
\end{equation}

This results allows us to understand that the generalized thermostat
associated to the ensemble (\ref{gce}) can be taken without any lost of
generality when $N$ is large enough as \textit{an ordinary Gibbs thermostat
with a variable inverse temperature} (\ref{variable.b}) which fluctuates
around the equilibrium value $\beta^{\ast}=\beta\left(  E^{\ast};\eta\right)
$, being $E^{\ast}$ the equilibrium energy corresponding to the stationary
point $\Theta^{\ast}=\Theta\left(  E^{\ast}\right)  $. The inverse temperature
fluctuations of such generalized thermostat $\delta\beta=\beta\left(
E;\eta\right)  -\beta\left(  E^{\ast};\eta\right)  \simeq\left(  \eta
\partial^{2}\Theta\left(  E^{\ast}\right)  /\partial E^{2}\right)  \ast\delta
E$ are correlated to the energy fluctuations of the interest system $\delta
E=E-E^{\ast}$ as follows:%
\begin{equation}
\left\langle \delta\beta\delta E\right\rangle \simeq\eta\frac{\partial
^{2}\Theta\left(  E^{\ast}\right)  }{\partial E^{2}}\left\langle \delta
E^{2}\right\rangle .
\end{equation}
Taking into account the approximation $\delta\Theta=\Theta\left(  E\right)
-\Theta\left(  E^{\ast}\right)  \simeq\left[  \partial\Theta\left(  E^{\ast
}\right)  /\partial E\right]  \ast\delta E$ and the well-known relation
between the fluctuations $\left\langle \delta\Theta^{2}\right\rangle $ with
the entropy Hessian $\partial^{2}S/\partial\Theta^{2}$ within the Gaussian
approximation $\left\langle \delta\Theta^{2}\right\rangle \simeq-\left(
\partial^{2}S\left(  \Theta^{\ast}\right)  /\partial\Theta^{2}\right)  ^{-1}$
(now in terms of the energy reparametrization $\Theta$), the inverse of the
average square dispersion of the energy $\left\langle \delta E^{2}%
\right\rangle $ can be expressed as follows:%
\begin{equation}
\left\langle \delta E^{2}\right\rangle ^{-1}=-\left(  \frac{\partial
\Theta\left(  E^{\ast}\right)  }{\partial E}\right)  ^{2}\frac{\partial
^{2}S\left(  \Theta^{\ast}\right)  }{\partial\Theta^{2}}. \label{dE2}%
\end{equation}
Finally, by combining the stationary condition (\ref{station}) and the
transformation rule (\ref{transf}):%
\begin{equation}
\eta\frac{\partial^{2}\Theta}{\partial E^{2}}\equiv\frac{\partial S}%
{\partial\Theta}\frac{\partial^{2}\Theta}{\partial E^{2}}=\frac{\partial^{2}%
S}{\partial E^{2}}-\left(  \frac{\partial\Theta}{\partial E}\right)  ^{2}%
\frac{\partial^{2}S}{\partial\Theta^{2}},
\end{equation}
as well as the equation (\ref{dE2}), the thermodynamic uncertainty relation
(\ref{tur}) is obtained.

Let us now consider that the interest system becomes extensive with the
imposition of the thermodynamic limit and the same one exhibits an ensemble
inequivalence inside the energy interval $\left(  \varepsilon_{1}%
,\varepsilon_{2}\right)  $, where $\partial^{2}s\left(  \varepsilon
_{i}\right)  /\partial\varepsilon^{2}<0$, being $\varepsilon=E/N$ and $s=S/N$
the energy and the entropy per particle respectively. Everything that should
be taking into consideration for the convergence of the Metropolis algorithm
described above is the using of an appropriate energy reparametrization
$\Theta=\Theta\left(  E\right)  $ \cite{vel-mmc}. Obviously, there exist an
undeterminable number of possibilities in order to perform such selection.

A convenient choice of the energy reparametrization under the above background
conditions is given by $\Theta\left(  E\right)  =N\varphi\left(  E/N\right)
$, where the first derivative of the bijective function $\varphi\left(
\varepsilon\right)  $ is taken as follows:%
\begin{equation}
\xi\left(  \varepsilon\right)  =\frac{\partial\varphi\left(  \varepsilon
\right)  }{\partial\varepsilon}=\exp\left[  -\lambda\left(  \varepsilon
_{2}-\varepsilon\right)  \right]  \label{chita}%
\end{equation}
being $\lambda$ a large enough positive constant \cite{vel-mmc}. Thus, the
effective inverse temperature of the thermostat in terms of the instantaneous
value of the energy per particle $\varepsilon$ of the interest system is given
by $\beta\left(  \varepsilon;\eta\right)  =\eta\xi\left(  \varepsilon\right)
$. It can be verified that an unnecessary large value of $\lambda$ leads to an
increasing of the inverse temperature fluctuations. An appropriate
prescription for the value of $\lambda$ is given by $\lambda\gtrapprox
\beta_{c}^{-1}$, being $\beta_{c}$ an estimation of the critical inverse
temperature of the first-order phase transition. Each point of the caloric
curve $\beta$ \textit{versus} $\varepsilon$ is determined from the averages
$\left\langle \varepsilon\right\rangle $ and $\left\langle \beta\left(
\varepsilon;\eta\right)  \right\rangle $ taken from a Metropolis dynamics by
keeping fixed the canonical parameter $\eta$ of the generalized thermostat.
The second derivative of the entropy per particle is obtained from the
thermodynamic uncertainty relation (\ref{tur}). The whole simulation can start
from $\eta=\beta_{2}$, being $\beta_{2}$ the corresponding inverse temperature
at the point $\varepsilon_{2}$, where $\eta$ is increased until the caloric
curve be obtained for all the energetic interval $\left(  \varepsilon
_{1},\varepsilon_{2}\right)  $.

The variant of the Metropolis algorithm described above seems to be a very
general alternative to avoid the anomalies related with the presence of a
first-order phase transitions in the short-range interacting systems. An
example of application was presented in the ref.\cite{vel-mmc} in order to
obtain the microcanonical caloric curve of the $q=10$ states Potts model
\cite{pottsm}:%
\begin{equation}
\hat{H}_{N}=\sum_{\left\{  ij\right\}  }\left(  1-\delta_{\sigma_{i}%
,\sigma_{j}}\right)  , \label{potts}%
\end{equation}
on a square lattice with $N=L\times L$ with periodic boundary conditions,
whose microcanonical description shows the existence of a first-order phase
transition. The sum is over pairs of nearest neighbor lattice points only and
$\sigma_{i}$ is the spin state at the \textit{i-th} lattice point.

Generally speaking, any Metropolis algorithm is more inefficient than other
nonlocal Monte Carlo methods involving clusters algorithms. However, the use
of such clusters algorithms can be limited by the phenomenon of the ensemble
inequivalence. For example, the clusters algorithms for the Potts model
\cite{wolf} are based on a mapping of this model system to a random clusters
model of percolation throughout the equation:
\begin{equation}
Z=\sum_{\left\{  \sigma\right\}  }\exp\left[  -\beta\hat{H}_{N}\right]
=\sum_{\left\{  n\right\}  }q^{N_{c}}p^{b}\left(  1-p\right)  ^{Nd-b}%
,\label{cluster}%
\end{equation}
where $p=1-e^{-\beta}$, $b$ is the number of bonds, $N_{c}$ the number of
clusters and $d$ the dimension of the lattice. Whilst such Monte Carlo methods
allow in general an efficient calculation of the thermodynamic properties of
the Potts models, they can undergo a supercritical slowing down whenever an
ensemble inequivalence exists within the canonical description.

The ensemble inequivalence can be avoided by using the generalized canonical
description (\ref{gce}) instead of the usual canonical description. A very
important question is how to develop suitable clusters algorithms within the
framework of the generalized canonical ensemble (\ref{gce}). A serious
obstacle is the nonlinear character of the energy reparametrization
$\Theta=\Theta\left(  E\right)  $, i.e.: the linear character of the energy
reparametrization $\Theta\left(  E\right)  =E$ is crucial for the mapping of
the Potts model described by the equation (\ref{cluster}). However, there
could be a very simple way to avoid such difficulties. The reader can notice
that the thermodynamic uncertainty relation (\ref{tur}) does not make an
explicit reference to the particular energy reparametrization $\Theta\left(
E\right)  $ used during its derivation. This remarkable feature suggests us
that actually there is not needing to perform an exact Monte Carlo
implementation of the generalized canonical ensemble\textit{ }(\ref{gce}) when
the system size $N$ is large enough, that is,\textit{ }all that it needs to
demand in order to avoid the ensemble inequivalence is the using of a
generalized thermostat with a fluctuating inverse temperature.%

\begin{figure}
[t]
\begin{center}
\includegraphics[
height=2.8106in,
width=3.5129in
]%
{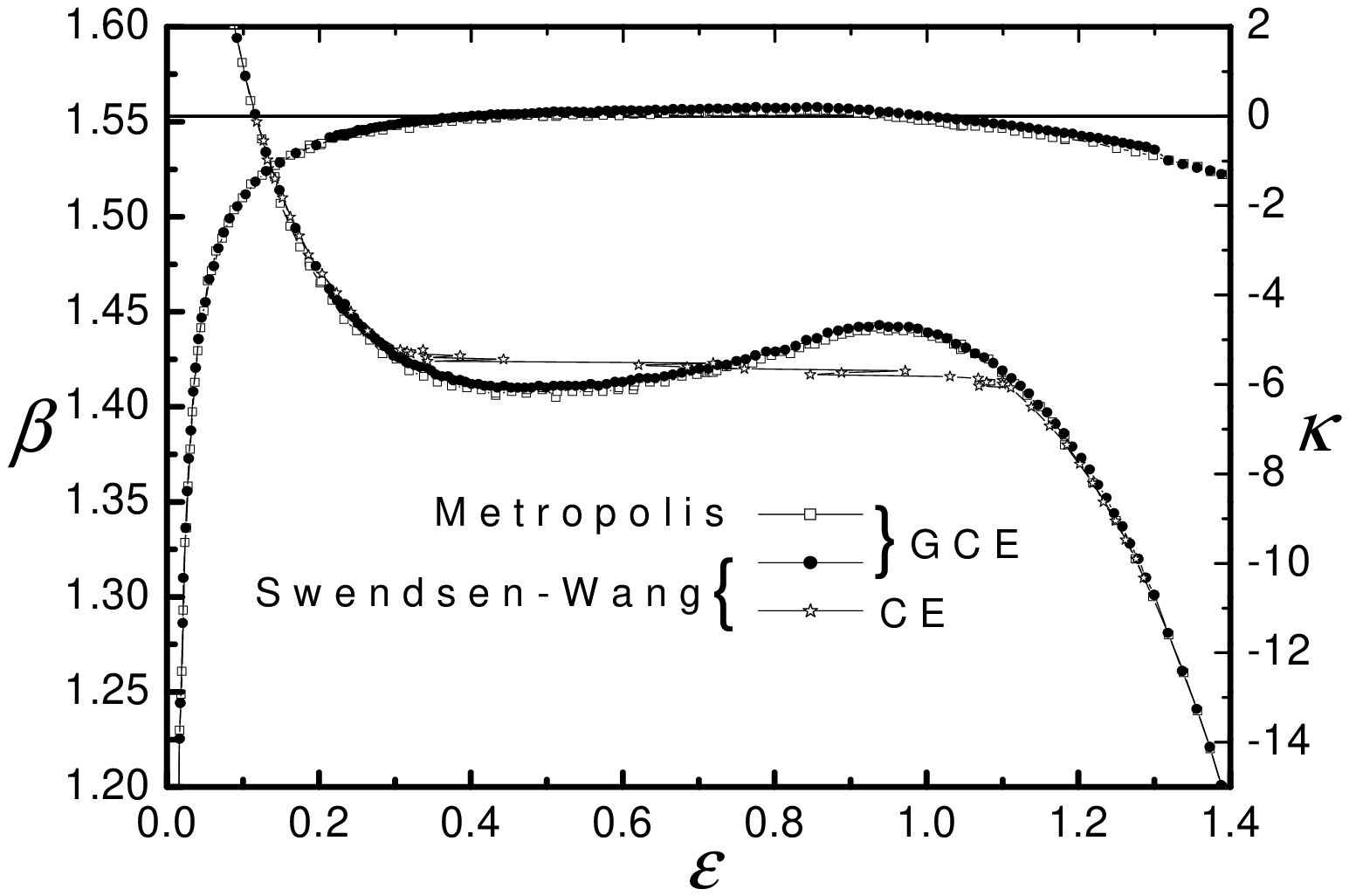}%
\caption{Caloric $\beta\left(  \varepsilon\right)  $ and curvature
$\kappa\left(  \varepsilon\right)  $ curves obtained from the Metropolis
algorithm and the Swendsen-Wang cluster algorithm using thermostats associated
to the generalized canonical ensemble (GCE) as well as the canonical ensemble
(CE) by considering $10^{5}$ iterations for each point. }%
\label{compara.eps}%
\end{center}
\end{figure}

Such a working hypothesis leads to propose the following scheme for enhancing
the potentialities of any Monte Carlo algorithm based on the Gibbs canonical
ensemble in order to deal with the ensemble inequivalence: (1)\textit{ }To
generate a new configuration $X_{i}$ from the previous configuration $X_{i-1}$
by using a Monte Carlo method with constant inverse temperature $\beta$; (2)
Redefine the inverse temperature of the thermostat $\beta=\beta\left(
E;\eta\right)  $ of the next configuration by using the energy $E=H_{N}\left(
X_{i}\right)  $ of the present configuration $X_{i}$. The function
$\beta\left(  E;\eta\right)  $ can be taken as the effective inverse
temperature (\ref{variable.b}) associated to the Metropolis algorithm based on
the generalized canonical ensemble (\ref{gce}). The microcanonical caloric
curve $\beta\left(  \varepsilon\right)  =\partial s\left(  \varepsilon\right)
/\partial\varepsilon$ can be obtained from the average values of the energy
$\left\langle \varepsilon\right\rangle $\ and the inverse temperature
$\left\langle \beta\right\rangle =\eta\left\langle \xi\left(  \varepsilon
\right)  \right\rangle $, while the curvature or the second derivative of the
entropy per particle $\kappa\left(  \varepsilon\right)  =\partial^{2}s\left(
\varepsilon\right)  /\partial\varepsilon^{2}$ by means of the relation
$\kappa\left(  \varepsilon\right)  =\left(  \sigma_{\beta}\sigma_{\varepsilon
}-1\right)  /\sigma_{\varepsilon}^{2}$, being $\sigma_{\varepsilon}%
^{2}=N\left\langle \delta\varepsilon^{2}\right\rangle $, $\sigma_{\beta}%
^{2}=N\left\langle \delta\beta^{2}\right\rangle $ and $\xi\left(
\varepsilon\right)  $ the function (\ref{chita}).

A complete analysis about the\ applicability and the convergence of the
procedure proposed above to any Monte Carlo algorithm based on the Gibbs
canonical ensemble seems to be at a first glance a very difficult mathematical
task. Nevertheless, the thermodynamic relation (\ref{tur}) constitutes by
itself a general result supporting in principle the applicability of such
methodology. 

Numerical evidences seems to be in agreement with the present viewpoint. The
FIG.\ref{compara.eps} shows a study where the above procedure was employed in
order to combine the potentialities of the well-known Swendsen-Wang cluster
algorithm (SW) \cite{wolf} and the use of a generalized thermostat with a
fluctuating inverse temperature in order to access to the microstates with a
negative heat capacity presented in the thermodynamical description of the
$q=10$ states Potts model (\ref{potts}). A previous simulation (by using the
usual Metropolis or SW algorithms) allows to set the interest energetic window
between $\varepsilon_{1}=0.2$ and $\varepsilon_{2}=1.2$ which encloses the
region of ensemble inequivalence of this model for $L=25$. The inverse
critical temperature was estimated as $\beta_{c}\simeq1.4$, allowing us to set
$\lambda=0.8$. The reader can notice that while the SW algorithm with a
constant temperature is unable to reproduce the backbending behavior of the
microcanonical caloric curve, the use of a thermostat with a fluctuating
inverse temperature avoids such limitation both for this cluster algorithm as
well as the Metropolis one. The agreement between these last Monte Carlo
methods is remarkable.

Thus, the analysis of the thermodynamic uncertainty relation (\ref{tur})
consequences opens a whole world of theoretical developments and numerical
studies about the enhancing of the available Monte Carlo methods based on the
canonical ensemble in order to deal with the ensemble inequivalence phenomenon
without appealing to the multicanonical methods \cite{mc3}.


\begin{thebibliography}{9}                                                                                                %
\bibitem {mc3}P. D. Landau and K. Binder, \textit{A guide to Monte Carlo
simulations in Statistical Physics} (Cambridge Univ Press, 2000).

\bibitem {vel.geo}L. Velazquez and F. Guzmann, e-print (2006)
[cond-mat/0610712]. L. Velazquez, e-print (2006) [cond-mat/0611595].

\bibitem {vel-mmc}L. Velazquez and J. C. Castro Palacio,\ e-print (2006) [cond-mat/0606727].

\bibitem {gro1}D.H.E Gross, \textit{Microcanonical thermodynamics: Phase
transitions in Small systems}, \textit{66 Lectures Notes in Physics}, (World
scientific, Singapore, 2001).

\bibitem {jaynes}E. T. Jaynes, Phys. Rev. \textbf{106} (1957) 620.

\bibitem {toral1}R. Toral, Physica A \textbf{365} (2006) 85.

\bibitem {metro}N. Metropolis, A. W. Rosenbluth, M. N. Rosenbluth, A. H.
Teller and E. Teller, J. Chem. Phys. \textbf{21} (1953) 1087.

\bibitem {pottsm}J. -S. Wang, R. H. Swendsen and R. Koteck\'{y}, Phys. Rev.
Lett. \textbf{63} (1989) 1009.

\bibitem {wolf}U. Wolff, Phys. Rev. Lett. \textbf{62} (1989) 361.
\end{thebibliography}
\end{document}